\documentclass[journal]{IEEEtran}
\ifCLASSINFOpdf
\else
\fi
\usepackage{cite}

\usepackage{amsmath,amssymb,amsfonts}
\usepackage{breqn}

\usepackage{graphicx}
\graphicspath{{Images/}}

\usepackage{booktabs}
\usepackage{multirow}
\usepackage{tabularx,multirow,ragged2e}
\newcolumntype{Y}{>{\RaggedRight\arraybackslash}X}
\newcolumntype{P}[1]{>{\RaggedRight\arraybackslash}p{#1}}

\usepackage[utf8]{inputenc}
\usepackage{dirtytalk}
\usepackage{tabu}
\usepackage{subfigure}
\usepackage{color}
\usepackage{comment}
\usepackage{xspace}
\usepackage{booktabs}
\usepackage{float}

\usepackage{subfigure}

\usepackage{url}

\usepackage{nomencl}
\makenomenclature

\usepackage{ifthen}
  \renewcommand{\nomgroup}[1]{%
  \item[\bfseries
  \ifthenelse{\equal{#1}{P}}{Sets, Parameters}{%
  \ifthenelse{\equal{#1}{V}}{Variables}{}}%
  ]}

\allowdisplaybreaks

\begin{document}
%
\title{Achieving Disaster-Resilient Distribution Systems via Emergency Response Resources: A Practical Approach}



%


\author{Santosh~Sharma,~\IEEEmembership{Student~Member,~IEEE,}
       Qifeng~Li,~\IEEEmembership{Member,~IEEE,} Qiuhua~Huang,~\IEEEmembership{Member,~IEEE,}
        and~Ahmad~Tbaileh,~\IEEEmembership{Member,~IEEE}
\vspace{-8mm}\thanks{This work is supported by the U.S. Department of Energy under Contract DE-AC05-76RL01830 ({\it Corresponding author: Qifeng Li}).}
\thanks{S. Sharma and Q. Li are with the Department of Electrical Engineering $\&$ Computer Science, University of Central Florida. Q. Huang and A. Tbaileh are with the Electricity Security Group, Pacific Northwest National Laboratory.}
}


\maketitle 

\begin{abstract}

This paper presents a practical approach to utilizing emergency response resources (ERRs) and post-disaster available distributed energy resources (PDA-DERs) to improve the resilience of power distribution systems against natural disasters. The proposed approach consists of two sequential steps: first, the minimum amount of ERRs is determined in a pre-disaster planning model; second, a post-disaster restoration model is proposed to co-optimize the dispatch of pre-planned ERRs and PDA-DERs to minimize the impact of disasters on customers, i.e., unserved energy for the entire restoration window. Compared with existing restoration strategies using ERRs, the proposed approach is more tractable since 1) in the pre-disaster stage, the needed EERs are determined based on the prediction of energy shortage and disaster-induced damages using machine learning-based algorithms (i.e., cost-sensitive-RF-QRF for prediction of outage customers, random forest for prediction of outage duration, and CART for prediction of disaster-induced damages); 2) in the post-disaster stage, the super-node approximation (SNA) and the convex hull relaxation (CHR) of distribution networks are introduced to achieve the best trade-off between computational burden and accuracy. Tests of the proposed approach on IEEE test feeders demonstrated that a combination of SNA and CHR remarkably reduces the solution time of the post-disaster restoration model.

\end{abstract}


\vspace{-3mm}


\nomenclature[P]{$C_{j}^{MDG},C_{j}^{MES},C_{j}^{MPV}$}{Per unit cost of MDGs, MESSs, MPVs of type $j$}
\nomenclature[V]{$N_{i,j}^{z}, z \in \Omega_{mers}$}{Required number of MERs (MDGs, MPVs, MESSs) of size type $j$ at node $i$}
\nomenclature[V]{$N_{j}^{g}, g \in \Omega_{mers}$}{Total number of required MERs (MDGs, MPVs, MESSs) of size type $j$}
\nomenclature[P]{$p_{j}^{MDG\_s},p_{j}^{MPV\_s}$}{Size (MW) of MDGs, MPVs of size type $j$}
\nomenclature[P]{$E_{j}^{MES\_s},S_{j}^{MES\_s}$}{Size (MWh, MVA) of MESSs of size type $j$}
\nomenclature[V]{$p_{i,t}^{x},q_{i,t}^{x}, x \in \Omega_{mers}$}{Active \& reactive power output of MERs at node $i$ at time $t$}
\nomenclature[V]{$p_{i,t}^{x},q_{i,t}^{x}, x \in \Omega_{ders}$}{Active \& reactive power output of DERs at node $i$ at time $t$}
\nomenclature[P]{$p_{i}^{DG\_t},q_{i}^{DG\_t}$}{Total active, reactive power capacity of DGs at node $i$}
\nomenclature[P]{$E_{i}^{ES\_t},S_{i}^{ES\_t}$}{Total MWh, MVA capacity of ESSs at node $i$}
\nomenclature[P]{$p_{i,t}^{PV\_t}$}{Active power capacity of PVs at node $i$ at time $t$}
\nomenclature[P]{$\overline{p_{i,t}^{L}},\underline{p_{i,t}^{L}},\overline{q_{i,t}^{L}},\underline{q_{i,t}^{L}}$}{Total and critical active and reactive power demand at node $i$ at time $t$}
\nomenclature[V]{$v_{i,t}$}{Squared of voltage at node $i$ at time $t$}
\nomenclature[P]{$r_{ik}, x_{ik}$}{Resistance and reactance of line $ik$}
\nomenclature[V]{$p_{ik,t}, q_{ik,t}$}{Active and reactive power flow on line $ik$ at time $t$}
\nomenclature[V]{$\ell _{ik,t}$}{Squared of current flow on line $ik$ at time $t$}
\nomenclature[P]{$p_{i,t}^{G},q_{i,t}^{G}$}{Active, reactive grid power at node $i$ at time $t$}
\nomenclature[P]{$\underline{v_{i}}, \overline{v_{i}}$}{Minimum and maximum limits of voltage at node $i$}
\nomenclature[P]{$\overline{\ell _{ik}}$}{Squared of current capacity limit of line $ik$}
\nomenclature[P]{$\Omega_{ders},\Omega_{mers}$}{Set of DERs \& MERs}
\nomenclature[P]{$\Omega_{res}$}{Set of MDGs, MPVs, DGs, PVs}
\nomenclature[P]{$\Omega_{s}$}{Set of MESSs \& ESSs}
\nomenclature[V]{$u_{ik,t}$}{Operating status of line $ik$ at time $t$}
\nomenclature[V]{$P^{y,l}_{i,t}, y \in \Omega_{s}$}{Power loss in energy storage at node $i$ at time $t$}
\nomenclature[P]{$E^{y,spl}_{i}, y \in \Omega_{s}$}{Surplus energy in energy storage at node $i$}
\nomenclature[P]{$r^{c,e}_{i},r^{c,ct}_{i}, c \in \Omega_{s}$}{Total \& converter resistance in energy storage at node $i$}
\nomenclature[P]{$M$}{A scalar constant}
\nomenclature[P]{$\overline{S_{ik}}$}{Power carrying capacity limit of line $ik$}
 
\printnomenclature[1in]

%
\IEEEpeerreviewmaketitle

\vspace{0mm}
\section{Introduction} \label{sec:introduction}

\IEEEPARstart{I}{n} recent years, high-impact low-frequency (HILF) events such as hurricanes, ice storms, earthquakes, cyber attacks, etc. are happening at a higher frequency \cite{hines2008trends}. Impacts of such HILF events are colossal \cite{executive2013economic}, and it has been reported that such events may cause loss of billions of dollars to United States every year \cite{smith2013us}. An important measure to mitigate this issue is improving the resilience of critical infrastructure (CI) systems such as electricity, water delivery, transportation, communication systems, health, finance etc. Among all, the power system plays a fundamental role since all other CI systems rely heavily on electricity. There is no commonly accepted definition for resilience of power systems thus far. According to \cite{house2013presidential}, power system resilience can be considered as \say{the ability to prepare for and adapt to changing conditions and withstand and recover rapidly from extreme outages.} Therefore, strategies for fast and effective restoration of power supply after extreme events play an important role in power system resilience.

Early-stage research activities related to power system restoration are mainly focused on the black-start of transmission and generation systems \cite{nagata2002multi, adibi2000power}. They have described restoration strategies and issues for bulk power system restoration using centralized generation and bulk transmission systems. Utilization of PDA-DERs along with microgrid and networked microgrids formation after natural hazards was studied in the last few years \cite{arif2017networked, Chen2016}. Here, the PDA-DERs are referred to as the grid-connected DERs that survive from the disaster, for example, diesel generators (DGs), energy storage systems (ESSs), and photovoltaics (PVs). Recently, some researchers started exploring the possibility of using ERRs, such as repair crews (RCs) and mobile energy resources (MERs) \cite{kim2018enhancing,lei2016mobile}, to accelerate the power systems recovery. Existing research results show that this is an effective strategy with high potential.  

However, there are still some critical questions to be answered. For instance, how to properly determine the needed ERRs before the disaster? A scenario-based method for the pre-allocation and pre-positioning of MERs is investigated in \cite{kim2018enhancing,lei2016mobile}. However, the huge number of scenarios generated to simulate the post-disaster conditions makes the proposed methods not practical in real-world applications. Moreover, which type, what size of MERs, and which locations are the best for disaster-relief? These questions have not yet been well-studied in the literature.

In modern power systems, the penetration of grid-integrated distributed energy resources is undergoing a rapid growth \cite{li2016convex}. In \cite{arif2017networked, Chen2016}, the utilization of PDA-DERs with microgrid and networked microgrids formation has been shown as an effective strategy for distribution system recovery after natural disasters. However, many DERs are intermittent or non-dispatchable, which raises an important question: how to coordinate ERRs with PDA-DERs to expedite the post-disaster restoration.

Since the power system is in a state of chaos after disasters \cite{sharma2019scenario}, solving post-disaster restoration problems is computationally challenging, as reported in the literature. The main challenges arise from 1) nonlinearity of the power flow model, 2) nonlinearity of DERs and MERs models, and 3) combinatorial nature of the optimization problem. It is an urgent need to develop a computationally-effective solution method to solve the restoration problem of disaster-impacted distribution systems. 

\begin{figure}[!ht]
\centering
\includegraphics[scale=1]{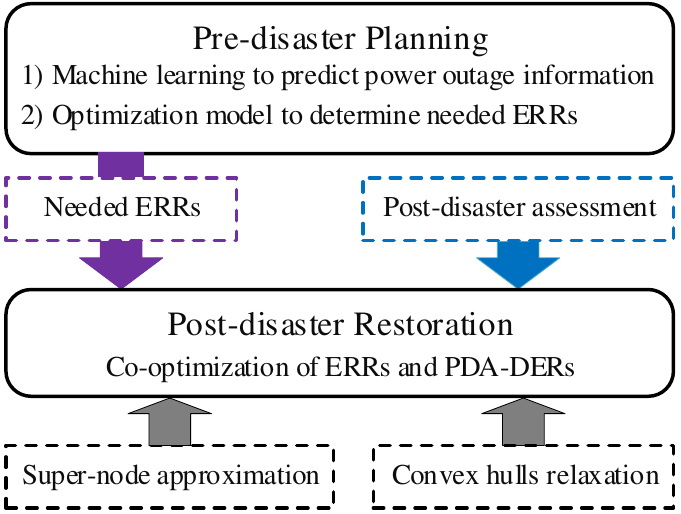}
\caption{The proposed framework for distribution system restoration.}
\label{Framework}
\end{figure}

This paper seeks answers to the above-mentioned critical questions, based on which a practical two-stage framework of utilizing ERRs and PDA-DERs to accelerate the restoration process of distribution systems after disasters, as shown in Figure \ref{Framework}, is developed. Note that power outage information collectively refers to post-disaster energy shortage and disaster-induced damages in Figure \ref{Framework}. Main contributions of this paper can be summarized as followings:
\begin{enumerate}
  \item In the pre-disaster stage, two-stage cost-sensitive-random forest-quantile regression forest (cost-sensitive-RF-QRF) for the prediction of outage customers \cite{kabir2019predicting}, classification and regression trees (CART) for the prediction of damaged poles/lines and DERs \cite{guikema2010prestorm}, and random forest (RF) for the prediction of outage duration \cite{nateghi2014forecasting} are leveraged to estimate the post-disaster energy shortage and disaster-induced damages, based on which the needed number, type, and size of ERRs are determined by a proposed optimization model. In this paper, energy shortage refers to the total unmet power demand of customers due to outages resulting from a disaster. Compared with existing approaches that determine the needed ERRs based on a huge number of scenarios \cite{kim2018enhancing,lei2016mobile}, the proposed method is more technically feasible.
  \item To improve the computational efficiency of the post-disaster restoration model, the super-node approximation (SNA) is introduced, which reduces the problem size without significant loss of accuracy (refer to the proof in Section III-C for more details). Then, the convex hull relaxation (CHR) is adopted to convexify the overall optimization model to reduce the computational burden further. 
\end{enumerate}

The rest of the paper is organized as follows: Section II provides the description of the pre-disaster planning, and Section III describes the post-disaster restoration. Section IV provides case studies and results, and Section V concludes with the conclusion and the potential future research.

\section{Pre-Disaster Planning}\label{sec:pre-event planning}
\subsection{Machine Learning Models to Predict Outages and DERs Availability}
\begin{figure}[!ht]
\centering
\includegraphics[scale=0.65]{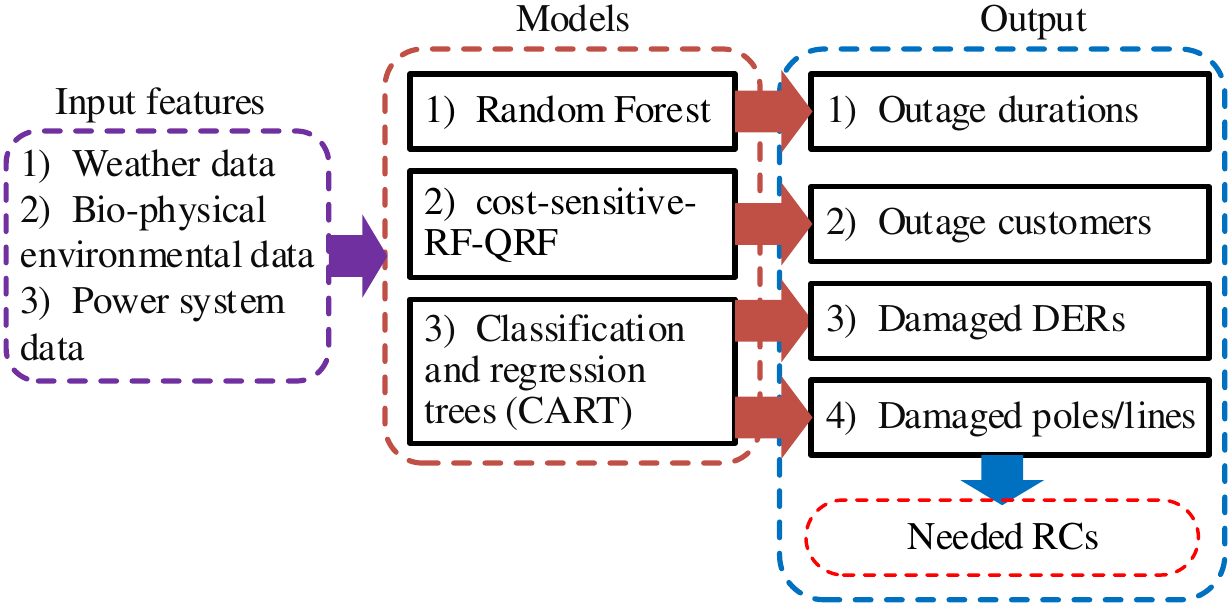}
\caption{Machine learning models to predict outage information.}
\label{MLmodels}
\end{figure}

To determine the needed ERRs for a geographical area in pre-disaster planning, post-disaster energy shortage and disaster-induced damages need to be predicted a period (generally a few days) before the disaster. The information about the post-disaster energy shortage includes the amount and duration of outage loads. Similarly, the information about disaster-induced damages refers to damaged poles/lines and DERs. There exist some research findings on the pre-disaster forecast of outage customers and durations, and damaged poles \cite{kabir2019predicting,nateghi2014forecasting,guikema2010prestorm}. In this paper, we adopt cost-sensitive-RF-QRF to predict the number of outage customers \cite{kabir2019predicting}, random forest (RF) to predict outage durations \cite{nateghi2014forecasting}, and CART to predict damaged poles/lines and DERs \cite{guikema2010prestorm}, as shown in Figure \ref{MLmodels}. The classification and regression trees (CART) based methods, which were proposed in \cite{guikema2010prestorm} to predict damaged poles, are modified to predict the damaged DERs in this paper. It is noted that outage prediction models adopted herein provide the aggregate forecasts of outage customers and durations, damaged poles/lines, and damaged DERs. Although the above-mentioned forecast models may not be perfectly accurate, we believe that they can provide necessary information for ERRs planning.


From the estimates of outage customers and duration and total connected loads of customers, we obtain total energy shortage (in MW and MVAr) and shortage duration (in hours) for a geographical area to get hit by a disaster. Similarly, from the predictions of damaged poles and lines, the required repair crews (RCs) are determined, as shown in Figure \ref{MLmodels}. Calculation of the required number of RCs based on the estimates of damaged poles/lines and available resource capacity of an RC is a straightforward problem and is not discussed further in the paper for the brevity. As shown in Figure 2, all three outage prediction models have input features related to weather data, bio-physical environmental data, and power system data. For further understanding of the models, interested readers are referred to \cite{kabir2019predicting,nateghi2014forecasting,guikema2010prestorm}.

\vspace{-3mm}

\subsection{Discussion on Different Types of MERs}
\begin{figure}[!ht]
\centering
\includegraphics[scale=1]{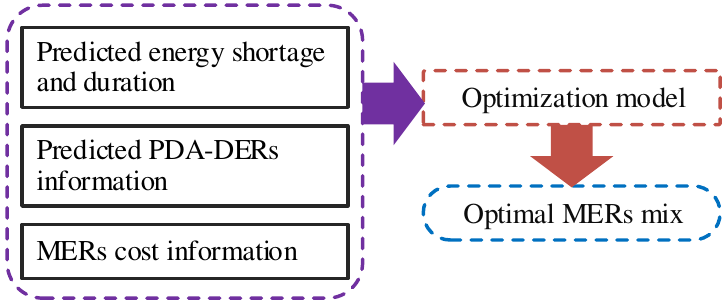}
\caption{Framework for optimal generation mix of of MERs.}
\label{TechnoEconomic}
\end{figure}

There exist various types of MERs, such as mobile diesel generators (MDGs), mobile energy storage systems (MESSs), and mobile photovoltaics (MPVs). The best type of MERs or the optimal mix of different types of MERs is determined based on the predicted PDA-DERs, costs of MERs, and predicted energy shortage and shortage durations, as shown in Figure \ref{TechnoEconomic}. MDGs have both operating and initial rental or purchasing costs while MESSs and MPVs do not have operating costs. Since the operating cost is generally much lower than the initial purchasing or rental cost, we do not consider the operating cost of MDGs in this paper. Generally, the initial purchasing or rental costs of MESSs and MPVs are high in comparison to MDGs. However, MESSs are desirable for a distribution grid with available renewable generators after a disaster. As such, in order to determine the optimal mix of different types of MERs for a specific post-disaster energy shortage, an optimal planning model of MERs will be designed in the next subsection based on the above discussion, as depicted in Figure \ref{TechnoEconomic}.

\vspace{-3mm}

\subsection{Optimization Model to Determine Needed MERs} \label{PDPO}
Using the above-predicted information of energy shortage and PDA-DERs as inputs, we design an optimization model to determine the needed MERs for a geographical area to get hit by a disaster. The objective function \eqref{Preobj} is chosen to minimize the total cost of MERs. The cost of MERs can be interpreted in different ways: the initial setup cost, or the actual purchasing, or rental cost. The decision variables of primary interest in the pre-disaster planning model are $N_{j}^{MDG}, N_{j}^{MES}, N_{j}^{MPV}$, which provide the optimal size, number, and type of MERs for a predicted post-disaster energy shortage. The overall optimization model is given as follows:
\begin{align} 
&\text{Min. }\left(\begin{array}{l} \sum\limits_{j=1}^{N_{j}^{MDG}} {C_{j}^{MDG}}{N_{j}^{MDG}} + \sum\limits_{j=1}^{N_{j}^{MES}}{C_{j}^{MES}}{N_{j}^{MES}} \\
+ \sum\limits_{j=1}^{N_{j}^{MPV}}{C_{j}^{MPV}}{N_{j}^{MPV}} \end{array} \right) \label{Preobj} \\
&\text{S.t.: }\nonumber \\
&p_{t}^{MDG}+p_{t}^{MES}+p_{t}^{MPV}+p_{t}^{DG}+p_{t}^{ES}+p_{t}^{PV}=p_{t}^{L} \label{Pblnc}\\
&q_{t}^{MDG}+q_{t}^{MES}+q_{t}^{MPV}+q_{t}^{DG}+q_{t}^{ES}+q_{t}^{PV}=q_{t}^{L} \label{Qblnc}\\
&\underline{\hat{p_{t}}^{L}}\leq p_{t}^{L}\leq \overline{\hat{p_{t}}^{L}} \label{Pload}\\
&\underline{\hat{q_{t}}^{L}}\leq q_{t}^{L}\leq \overline{\hat{q_{t}}^{L}} \label{Qload}\\
&0\leq p_{t}^{MDG}\leq \sum_{j}N_{j}^{MDG}p_{j}^{MDG\_s} \label{Pmdg}\\
&0\leq p_{t}^{DG}\leq \hat{p}^{DG\_t} \label{Pdg}\\
&p_{t}^{MPV} = \sum_{i}N_{i}^{MPV}p_{i}^{MPV\_s} \label{Pmpv}\\
&p_{t}^{PV} = \hat{p}_{t}^{PV\_t} \label{Ppv}\\
&k_{1}*p_{t}^{d}\leq q_{t}^{d}\leq k_{2}*p_{t}^{d}, d \in \Omega_{res} \label{Qres}\\
&0\leq E^{MES,spl}-\sum_{t}\left ( p_{t}^{MES}+p_{t}^{MES,l} \right )\Delta t
\nonumber\\
&\leq \sum_{j}N_{j}^{MES}E_{j}^{MES\_s} \label{Emess}\\
&\left ( p_{t}^{MES} \right )^2+\left ( q_{t}^{MES} \right )^2\leq \left( \sum_{j}S_{j}^{MES\_s}N_{j}^{MES} \right )^2 \label{Smess}\\
&0\leq \hat{E}^{ES,spl}-\sum_{t}\left( p_{t}^{ES}+p_{t}^{ES,l} \right ) \Delta t \leq \hat{E}^{ES\_t} \label{Eess}\\
&\left ( p_{t}^{ES} \right )^2+\left ( q_{t}^{ES} \right )^2\leq \left ( \hat{S}^{ES\_t} \right )^2, \label{Sess}
\end{align}
where equations \eqref{Pblnc} and \eqref{Qblnc} are power balance constraints, and inequalities \eqref{Pload} and \eqref{Qload} are load limits. Operating limits of MDGs, DGs, MPVs, and PVs are forced by inequalities \eqref{Pmdg}, \eqref{Pdg}, \eqref{Pmpv}, \eqref{Ppv}, and \eqref{Qres}. Note that an appropriate PV profile is used to modify the active power capacity of solar stations to limit power extraction only when solar power is available. To maintain near-constant power factor operation in machines and networks, constraint \eqref{Qres} ensures that the reactive power from MDGs, DGs, MPVs, and PVs is a fraction of respective active power output, where $k_{1}, k_{2}$ are constants with $k_{1} \leq k_{2} \leq 1$. Constraints \eqref{Emess} and \eqref{Eess} describe the state of charge (SoC) for MESSs and ESSs, respectively. Since power flow is not considered in the pre-disaster planning model, power loss terms in \eqref{Emess} and \eqref{Eess} are ignored; however, they are included in the post-disaster restoration where a post-disaster network is considered. Constraints \eqref{Smess} and \eqref{Sess} describe the MVA capacity limits of MESSs and ESSs. It should be noted that PVs are considered non-dispatchable; however, energy storage and DGs are considered dispatchable. Note that the parameters with a hat ( $\hat{}$ ) over them are the estimated parameters obtained from outage prediction models described in Section II-A. 

\section{Post-Disaster Restoration}
The different types of required MERs and RCs determined in Section \ref{sec:pre-event planning} will be prepared by the restoration engineers before the disaster. This section presents an optimization model to co-optimize 1) allocation and dispatch of pre-determined MERs in coordination with PDA-DERs to supply the outage loads, and 2) dispatch of pre-determined repair crews (RCs) to restore outage lines. Due to inherent nonlinearities and the combinatorial nature of the problem, the original optimization model is computationally intractable. For rapidly obtaining an accurate solution of this complex optimization problem, two techniques, i.e., convex-hull relaxation and super-node approximation, are introduced, which significantly improve the computational efficiency.  

\vspace{-3mm}

\subsection{Optimization Formulation}
In the restoration process, all available ERRs (including MERs and repair crews) and PDA-DERs should be optimally coordinated based on the network damage information obtained from post-disaster damage assessment. Choosing an objective function of minimizing the unserved energy, the post-disaster restoration model can be given as:   
\begin{align} 
&\text{Min. } \sum_{t} \sum_{i}\left ( \overline{p}_{i,t}^{L}- p_{i,t}^{L}\right )\Delta t \label{Pobj} \\
&\text{S.t.:}\nonumber \\
&\text{\eqref{Pload}-\eqref{Sess}}\nonumber \\
&-\left (1-u_{ik,t} \right )M\leq v_{i,t}-v_{k,t}-2\left ( r_{ik}p_{ik,t}+x_{ik}q_{ik,t} \right ) &\nonumber\\
&+\left ( \left ( r_{ik} \right )^{2}+\left ( x_{ik} \right )^{2}  \right )\ell _{ik,t}\leq \left (1-u_{ik,t} \right )M \label{Vol}\\
&\left ( p_{ik,t} \right )^{2}+\left ( q_{ik,t} \right )^{2}= v_{i,t}\ell _{ik,t}\label{flow}\\
&\left ( \underline{v}_{i} \right )^{2}\leq v_{i,t}\leq \left ( \overline{v}_{i} \right )^{2} \label{Vlimits}\\
&0\leq \ell _{ik,t}\leq \overline{\ell} _{ik} \label{Tlimits}\\
&\left ( p_{ik,t} \right )^{2}+\left ( q_{ik,t} \right )^{2}\leq \left ( \overline{S}_{ik} \right )^{2} \label{Slimits}\\
&r_{i}^{c,e}\left ( p_{i,t}^{c} \right )^2+r_{i}^{c,ct}\left ( q_{i,t}^{c} \right )^2=p_{i,t}^{c,l}v_{i,t}, c \in \Omega_{s} \label{Eloss}\\
&p_{i,t}^{G}+p_{i,t}^{MDG}+p_{i,t}^{MES}+p_{i,t}^{DG}+p_{i,t}^{ES}\nonumber\\
&+p_{i,t}^{PV}-p_{i,t}^{L}= \sum_{j}\left ( p_{ji,t}-r_{ji}\ell _{ji,t} \right )+\sum_{k}p_{ik,t} \label{BlncP}\\
&q_{i,t}^{G}+q_{i,t}^{MDG}+q_{i,t}^{MES}+q_{i,t}^{DG}+q_{i,t}^{ES}\nonumber\\
&+q_{i,t}^{PV}-q_{i,t}^{L}= \sum_{j}\left ( q_{ji,t}-x_{ji}\ell _{ji,t} \right )+\sum_{k}q_{ik,t} \label{BlncQ}\\
&\sum\limits_{i}N_{i,j}^{g} \leq N_{j}^{g}, \forall j, g \in \Omega_{mers}  \label{Nmer}\\
&-u_{ik,t}M\leq p_{ik,t}\leq u_{ik,t}M \label{Line1}\\
&-u_{ik,t}M\leq q_{ik,t}\leq u_{ik,t}M \label{Line2}\\
&u_{ik,t}\geq u_{ik,t-1} \label{Line3}\\
&\sum_{ik}\alpha^{m}_{ik,t}\leq 1, \forall m, \forall t \label{RC1}\\
&\alpha^{m}_{ji,t+\tau}+\alpha^{m}_{ik,t}\leq 1, \forall m, \forall \tau \leq TT_{ji,ik} \label{RC2}\\
&\beta_{ik,t}^{m}\leq \alpha_{ik,t}^{m} \label{RC3}\\
&\sum_{m} \beta^{m}_{ik,t}\leq 1, \forall ik, \forall t \label{RC4}\\
&u_{ik,t+1}\leq \frac{\sum_{m}\sum_{t}\beta_{ik,t}^{m}}{RT_{ik}}, \forall ik. \label{RC5}
\end{align}

The objective function \eqref{Pobj}, on the other hand, minimizes the restoration time of loads, since smaller energy not served for the entire horizon implies shorter restoration times of loads. Moreover, this objective function also expedites the restoration of outage lines as the faster the outage lines are restored, the quicker the loads are served, and the smaller the energy not served is. The decision variables of primary interest in post-disaster model are MER allocation variables ($N_{i,j}^{MDG}, N_{i,j}^{MES}, N_{i,j}^{MPV}$) and variable related to status of outage lines ($u_{ik,t}$).  

Constraints \eqref{Pload}-\eqref{Sess} in the pre-disaster model are also applicable in the post-disaster model. The difference is that the parameters with a hat ( $\hat{}$ ) over them are no longer estimated parameters by the outage prediction models; they are obtained from the post-disaster assessment here. Besides, the parameters and the variables in constraints \eqref{Pload}-\eqref{Sess} in the pre-disaster model are used to represent a whole geographical area. However, they are used to include all the islands formed after the disaster in the post-disaster model. Therefore, for example, $N_{j}^{MES}$ in \eqref{Emess} and \eqref{Smess} is replaced with $N_{i,j}^{MES}$ to allocate MERs in post-disaster islands.   

The \textit{DistFlow} model \cite{baran1989optimal,li2016convex} is adopted to model power flows \eqref{Vol} and \eqref{flow}. Voltage limits are imposed by the constraint \eqref{Vlimits}. Thermal and power carrying capacity of lines are forced by constraints \eqref{Tlimits} and \eqref{Slimits}, respectively. A high fidelity second-order model \cite{li2017convex} is adopted to model energy storage systems. The power loss due to charging and discharging in MESSs and ESSs is modeled by constraint \eqref{Eloss}. The nodal power balance is achieved via equations \eqref{BlncP} and \eqref{BlncQ}. Constraint \eqref{Nmer} ensures that the number of MERs in the post-disaster model does not exceed the number of MERs determined in the pre-disaster planning model. Constraints \eqref{Line1} and \eqref{Line2} are used to remove unavailable lines from the optimization model. Constraint \eqref{Line3} ascertains that once a line is operating, it remains operable.

Constraints \eqref{RC1}-\eqref{RC5} describe the dispatch of pre-determined repair crews (RCs). Note that constraints are applicable for both single or multiple repair crew dispatch. The binary variable $\alpha^{m}_{ik,t}$ provides the travelling status of RC $m$ in reference to line $ik$ at time $t$. Another binary variable, $\beta^{m}_{ik,t}$, is introduced, which provides the working status of RC $m$ in reference to line $ik$ at time $t$. If $\alpha^{m}_{ik,t}$ = 1, RC $m$ is at line $ik$ at time $t$. If $a^{m}_{ik,t}$ = 0, RC $m$ is either visiting other lines or traveling in the network at time $t$. Similarly, if $\beta^{m}_{ik,t}$ = 1, RC $m$ is repairing line $ik$ at time $t$. Constraint \eqref{RC1} represents that an RC can visit one line at a time. The constraint \eqref{RC2} means that an RC spends $TT_{ji,ik}$ time to travel from a line $ji$ to a line $ik$. The constraint \eqref{RC3} indicates that RC $m$ repairs a line $ik$ at time $t$ only if it is visiting the line $ik$ at time $t$. Constraint \eqref{RC4} indicates that only one repair crew can work at a line at a time; however, it is noted that it can be easily changed to allow multiple RCs to work on a line simultaneously. Constraint \eqref{RC5} represents that RCs should spend a minimum of $RT_{ik}$ time repairing a damaged line $ik$ before the status of the damaged line $ik$ can be changed to operating.

\vspace{-3mm}

\subsection{Convex Hull Relaxation}
The post-disaster restoration model presented in Section III-A is a mixed-integer nonlinear program (MINLP) problem, where the nonlinearities come from the power flow equation \eqref{flow} and the energy storage model \eqref{Eloss}. Such an MINLP form of a larger test system is computationally expensive. Moreover, the global optimality of a solution can not be guaranteed either. To mitigate these issues, the convex hull relaxation\cite{li2017convex}, which is considered one of the tightest convex relaxation of the \textit{DistFlow} model in radial networks, is used to relax the original MINLP problem into a mixed-integer convex hull program (MICHP) problem. It has been shown in \cite{li2018micro} that an MICHP problem is much more computationally-tractable than an MINLP problem of the same size. The convex hulls relaxation of nonlinear nonconvex constraints \eqref{flow} and \eqref{Eloss} are given in the following equations \eqref{CHflow} and \eqref{CHEloss}, respectively. Please refer to Appendix-A for the definition of auxiliary variables adopted herein.
\begin{align} 
&\left\{ {\begin{array}{c} {{{\left\| {\boldsymbol{A}{\boldsymbol{x}_{ik,t}}} \right\|}_2} - {\boldsymbol{b}^T}{\boldsymbol{x}_{ik,t}} \leq 0}\\ {\boldsymbol{c}_{ik}^T{\boldsymbol{x}_{ik,t}} - {d_{ik}} \leq 0} \end{array}} \right. \label{CHflow}\\
&\left\{ {\begin{array}{c} {{{\left\| {\boldsymbol{A}{\boldsymbol{y}_{i,t}^{\alpha}}} \right\|}_2} - {\boldsymbol{b}^T}{\boldsymbol{y}_{i,t}^{\alpha}} \leq 0}\\ {{{\left\| {{\boldsymbol{A}_i^{\alpha}}{\boldsymbol{y}_{i,t}^{\alpha}}} \right\|}_2} - {\boldsymbol{b}^T}{\boldsymbol{y}_{i,t}^{\alpha}} \leq {e_i^{\alpha}}}\\ {\left( \boldsymbol{ {c}}^{\alpha}_{i}\right)^T{\boldsymbol{y}_{i,t}^{\alpha}} - {d_i^{\alpha}} \leq 0, \alpha \in \Omega_{s}} \end{array}} \right. \label{CHEloss}
\end{align}

Replacing \eqref{flow} and \eqref{Eloss} with \eqref{CHflow} and \eqref{CHEloss}, respectively, the MINLP model is relaxed to a mixed-integer convex hull program (MICHP), which is computationally more manageable. However, with the MICHP formulation, the optimal restoration problem  for a relatively large distribution system like the IEEE 123-node test feeder is still challenging to solve. To further mitigate this computational efficiency issue, the super-node approximation is introduced in the following subsection.

\vspace{-3mm}

\subsection{Super-node Approximation for Dimension Reduction}
\begin{figure}[ht]
\centering
\includegraphics[scale=0.46]{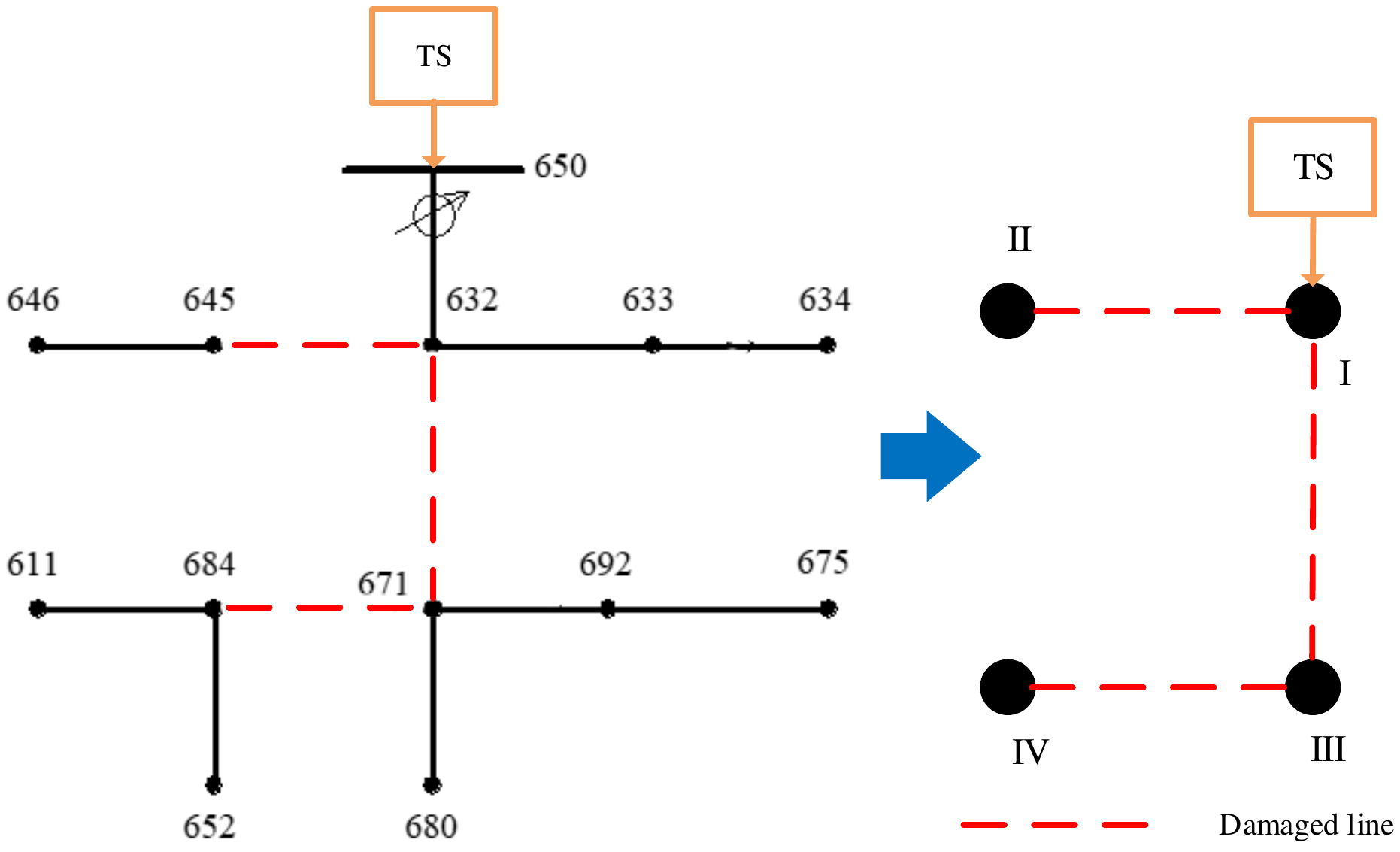}
\caption{Illustration of super-node approximation.}
\label{SNC12}
\end{figure}

The super-node approximation is introduced herein to reduce the dimension of large scale post-disaster systems, which, in turn, reduces the computational burden of the problem. A similar dimension reduction approximation referred to as the vertex collapse technique is used in graph theory \cite{ulusan2018restoration, hirschberg1979computing}. A disaster-impacted IEEE 13-node test feeder, shown in Figure \ref{SNC12}, is used to illustrate the super-node approximation in the context of distribution system restoration. After the disasters, the feeder is divided into four islands due to damage of some feeder lines. Each of these islands is represented by an aggregated node, which is referred to as a super-node in this paper. All PDA-DERs and loads within an island are aggregated at the corresponding super-node. For example, super-node I contains nodes 650, 632, 633, and 634, and their loads and PDA-DERs. Similarly, super-nodes II, III, IV represent the other three islands. As a result, the number of nodes considered in the computation is significantly reduced. For example, the IEEE 13 node test feeder with three damaged lines is converted to a network with four super-nodes. Therefore, instead of 13 nodes, four super-nodes are used in the optimization model, which is solved very efficiently, as shown in the next section.

\noindent
{\bf Remark.} {\it The super-node approximation can provide satisfactory accuracy in the context of post-disaster restoration in distribution systems.} 

To support the statement in the above remark, a mathematical analysis is given as follows: 
\subsubsection{Power losses in feeder lines are not of interest in the process of post-disaster restoration} Therefore, the terms in the DistFlow model that are related to losses can be ignored, which results in the following LinDistFlow model \cite{baran1989optimal,li2016convex}:
\begin{subequations}
\begin{align}
&p_{i,t}^{G}+p_{i,t}^{MDG}+p_{i,t}^{MES}+p_{i,t}^{DG}+p_{i,t}^{ES}  \nonumber\\
&+p_{i,t}^{PV}-p_{i,t}^{L}= \sum_{j}p_{ji,t}+\sum_{k}p_{ik,t} \label{activebalance} \\
&q_{i,t}^{G}+q_{i,t}^{MDG}+q_{i,t}^{MES}+q_{i,t}^{DG}+q_{i,t}^{ES} \nonumber\\
&+q_{i,t}^{PV}-q_{i,t}^{L}= \sum_{j} q_{ji,t}+\sum_{k}q_{ik,t} \label{reactivebalance}  \\
&v_{i,t}-v_{k,t}-2\left ( r_{ik}p_{ik,t}+x_{ik}q_{ik,t} \right )=0. \label{voltdiff}
\end{align}
\end{subequations}


\subsubsection{Voltage difference inside an electrical island is negligible} Note that a post-disaster electrical island is a part of a distribution network, which generally consists of a few nodes. Namely, the electrical distance between any two nodes inside an island is generally short with very small voltage difference. For example, consider nodes 692 and 675 in super-node III from figure \ref{SNC12}. Assume that node 692 is transmitting (4+j3) KVA of power to node 675 at 4.16 KV (1 p.u.). Using system parameters of the IEEE 13-node test feeder, the voltage of node 675 is calculated to be 4.03 KV, using equations \eqref{Vol} and \eqref{flow}. Hence, the voltage difference between node 692 and 675 is found to be 0.16 KV (0.03125 p.u.), which is a very small quantity and is negligible. Therefore, with the voltage difference omitted, constraint (\ref{voltdiff}) reduces to 
\begin{equation}
    v_{i,t}-v_{k,t} = 0 \label{voltdiff2},
\end{equation}
where $i$ and $k \in \mathcal{N}_j$ ($\mathcal{N}_j$ is the set of nodes of the $j$th island). Constraints (\ref{activebalance}), (\ref{reactivebalance}), and (\ref{voltdiff2}) together imply that all nodes in an island can be aggregated into one single node, i.e. the super-node. 

\begin{figure}[ht]
\centering
\includegraphics[scale=0.6]{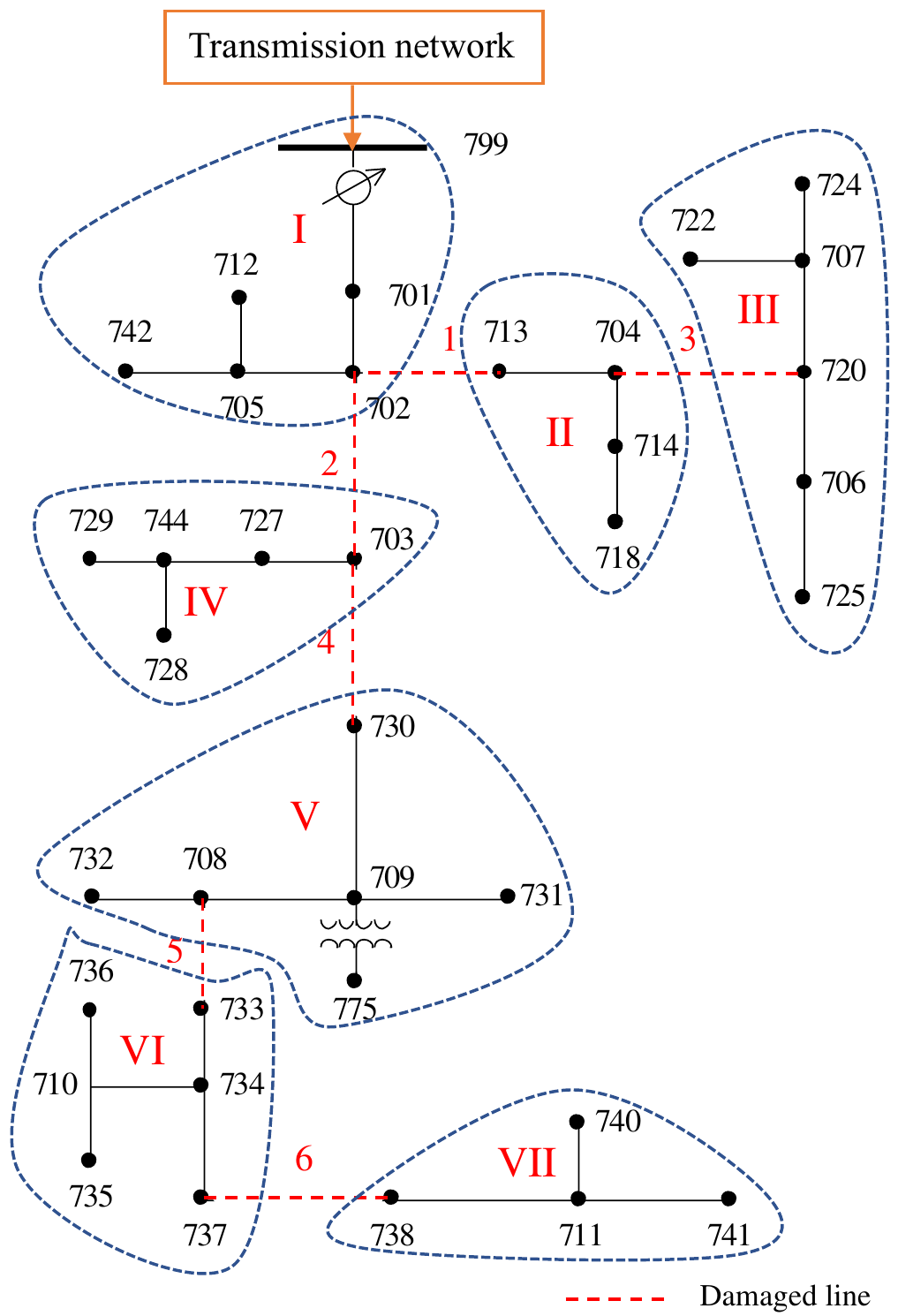}
\caption{Post-disaster distribution feeder for Case Study I.}
\label{CS1}
\end{figure}
\vspace{-3mm}

\begin{figure}[htbp]
\centering
\includegraphics[scale=0.535]{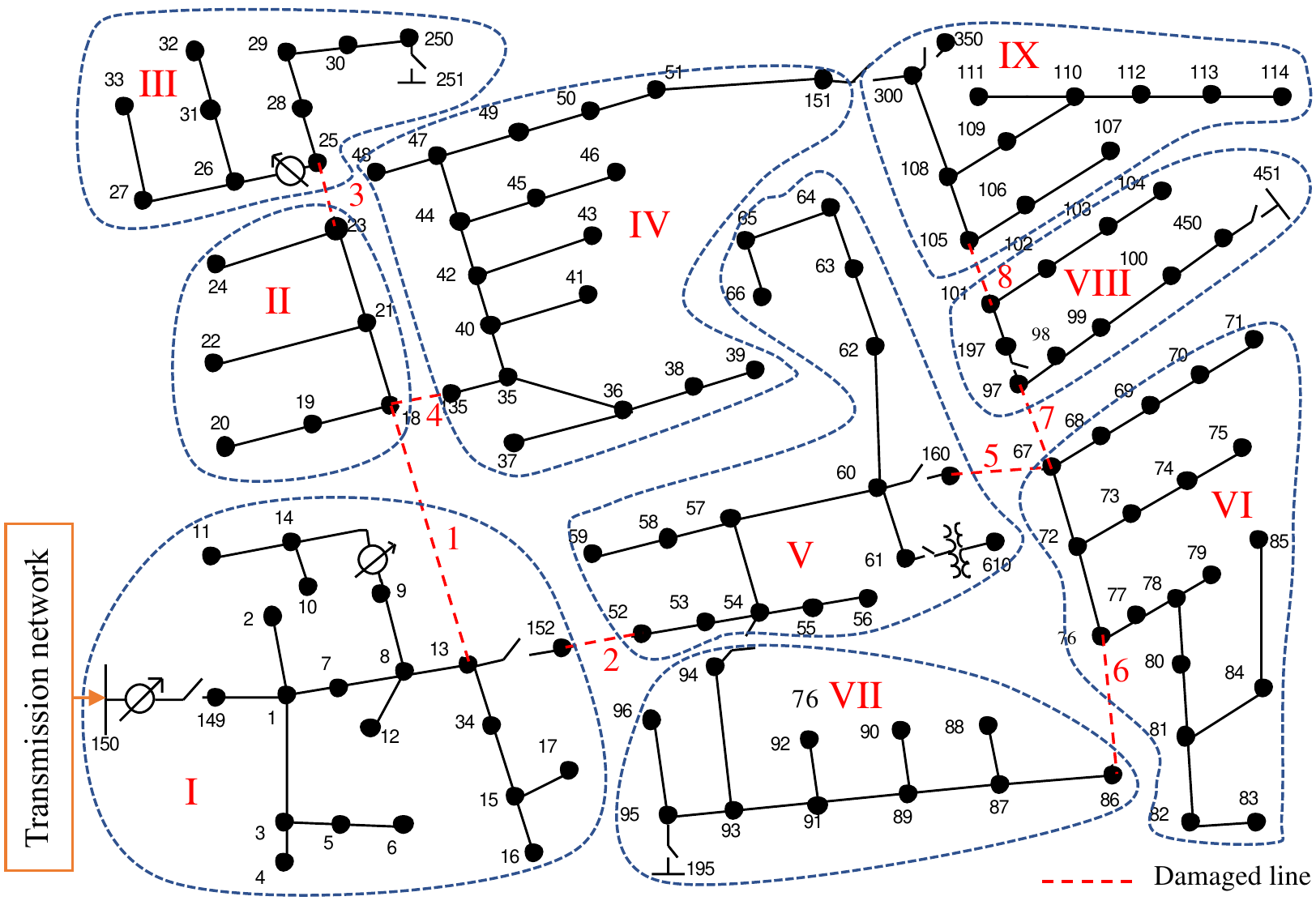}
\caption{Post-disaster distribution feeder for Case Study II.}
\label{CS2}
\end{figure}

\section{Case Study and Results} \label{sec:results}
This section presents the experimental results of testing the proposed approaches on the IEEE 37-node (Case Study I) and the IEEE 123-node (Case Study II) test feeders. The two test feeders are modified by replacing some lines with damaged lines to simulate the damages caused by the disaster, as shown in Figures \ref{CS1} and \ref{CS2}. For the 37-node test case, we assume that six lines are damaged in the disaster, and seven islands are formed after the disasters, as shown in Figure \ref{CS1}. For the 123-node test case, we assume that eight lines are damaged in the disaster, and nine islands are formed after the disaster, as shown in Figure \ref{CS2}. In real-world applications, such network outage information is obtained from the post-disaster damage assessment, which is not in the scope of this paper. The convex optimization problems were solved by GUROBI solver via the PYOMO package on a PC with a 64-bit Intel i5 dual-core CPU at 2.50 GHz and 16 GB of RAM while MINLP problems were solved by KNITRO solver using NEOS server.

\vspace{-3mm}

\begin{table}[htbp]
\centering
\caption{Predicted power shortage (Case Study I \& II).}
\label{table1}
\resizebox{\linewidth}{!}{%
\begin{tabular}{ccccccccc}
\hline \hline
Period              & 1  & 2  & 3  & 4  & 5  & 6  & 7  & 8  \\ \hline
Power shortage-CS: I (MW) & 2.21 & 2.46 & 2.33 & 2.09 & 2.09 & 2.21 & 2.46 & 2.33 \\ \hline 
Power shortage-CS: II (MW) & 3.14 & 3.49 & 3.32 & 2.97 & 2.97 & 3.14 & 3.49 & 3.32 \\ \hline \hline
\end{tabular}%
}
\end{table}

\vspace{-3mm}

\begin{table}[htbp]
\centering
\caption{Unit cost of available MER (Case Study I \& II).}
\label{table3}
\begin{tabular}{cccccc}
\hline \hline
\multicolumn{2}{c}{MDGs} & \multicolumn{2}{c}{MESSs} & \multicolumn{2}{c}{MPVs} \\ \hline
Size        & Cost       & Size            & Cost    & Size        & Cost       \\ \hline
1 MW      & \$1000     & 1.5 MWh/0.5 MVA  & \$1000  & 0.3 MW      & \$1000     \\
1.5 MW        & \$1500     & 2.5 MWh/1 MVA  & \$1500  & 0.4 MW        & \$1500     \\ \hline \hline
\end{tabular}%
\end{table}

\vspace{-3mm}

\begin{table}[htbp]
\centering
\caption{Locations of post-disaster available DERs (Case Study I \& II).}
\label{table2}
\resizebox{\linewidth}{!}{%
\begin{tabular}{ccccccc}
\hline \hline
                       & \multicolumn{2}{c}{DGs} & \multicolumn{2}{c}{ESSs}                              & \multicolumn{2}{c}{PVs} \\ \hline
                       & Capacity    & Node \#   & Capacity                        & Node \#             & Capacity    & Node \#   \\ \hline
CS-I                   & 200 KW      & 709       & 200 KWh/50 KVA                  & 720                 & 500 KW      & 701       \\ \hline
\multirow{2}{*}{CS-II} & 100 KW      & 94        & \multirow{2}{*}{300 KWh/75 KVA} & \multirow{2}{*}{25} & 2 MW        & 18        \\ \cline{2-3} \cline{6-7} 
                       & 200 KW      & 52        &                                 &                     & 2 MW        & 35        \\ \hline \hline
\end{tabular}%
}
\end{table}

\begin{table}[htbp]
\centering
\caption{Repair time of damaged lines (Case Study I \& II).}
\label{table5}
\begin{tabular}{ccccccccc}
\hline \hline
Line \#                  & 1 & 2 & 3 & 4 & 5 & 6 & 7 & 8 \\ \hline
Repair Time: CS-I (Hrs)  & 3 & 4 & 6 & 3 & 5 & 6 &   &   \\ \hline
Repair Time: CS-II (Hrs) & 4 & 3 & 5 & 4 & 7 & 2 & 5 & 7 \\ \hline \hline
\end{tabular}%
\end{table}

\begin{table}[htbp]
\centering
\caption{Optimal locations of needed MERs (Case Study I \& II).}
\label{table8}
\begin{tabular}{ccccc}
\hline \hline
      & \multicolumn{2}{c}{MDGs}      & \multicolumn{2}{c}{MESSs}     \\ \hline
      & Size          & Super-node \# & Size          & Super-node \# \\ \hline
CS-I  & 1.5 MW & I             & \multicolumn{2}{c}{}         \\ \hline
CS-II & 1 MW   & VI            & 2.5 MWh/1 MVA & I             \\ \hline \hline
\end{tabular}%
\end{table}

\begin{table}[htbp]
\centering
\caption{Computational performance of proposed solution approach.}
\label{tabletime}
\resizebox{\linewidth}{!}{%
\begin{tabular}{cccccc}
\hline \hline
                       &                 & \it{Base}          & \it{SNA}               & \it{CHR}    & \it{SNA+CHR}    \\ \hline
\multirow{2}{*}{CS-I}  & Objective value & N/A                 & N/A                 & 19.74 MWh & 19.72 MWh     \\ \cline{2-6} 
                       & Time            & \textgreater{}8h & \textgreater{}8h & $\approx$1h    & 33s  \\ \hline
\multirow{2}{*}{CS-II} & Objective value & N/A                 & N/A                 & 21.70 MWh  & 21.57 MWh      \\ \cline{2-6} 
                       & Time            & \textgreater{}8h & \textgreater{}8h & $\approx$6h    & 228s \\ \hline \hline
\end{tabular}%
}
\end{table}

\begin{table*}[htbp]
\centering
\caption{Operation time of outage lines (Case Study I \& II).}
\label{table7}
\begin{tabular}{ccccccccc}
\hline \hline
Line \#                 & 1 & 2 & 3 & 4 & 5 & 6 & 7 & 8 \\ \hline
Operation period: CS-I  & 2 (RC-2) & 2 (RC-1) & 4 (RC-1) & 5 (RC-2) & 4 (RC-2) & 7 (RC-1) &   &   \\ \hline
Operation period: CS-II & 2 (RC-1) & 2 (RC-2) & 7 (RC-1) & 8 (RC-2) & 6 (RC-2) & 3 (RC-1) & 4 (RC-2) & 5 (RC-1) \\ \hline \hline
\end{tabular}%
\end{table*}

\begin{figure}[htp]
  \centering
  \subfigure[Case Study I]{\includegraphics[scale=.85]{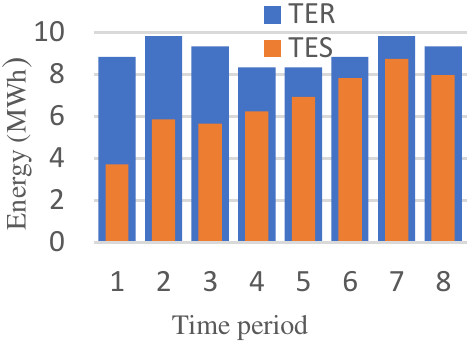}\label{ENS1}}\quad
  \subfigure[Case Study II]{\includegraphics[scale=.85]{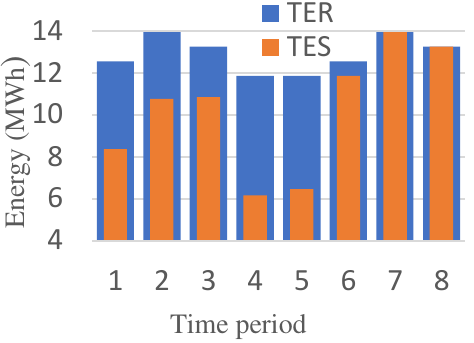}\label{ENS2}}
\caption{TER vs. TES comparison.}
\label{TERvsTES}
\end{figure}

Firstly, the pre-disaster planning model is executed to determine the optimal amount of ERRs needed in the post-disaster restoration and recovery based on the estimated energy shortage and shortage duration, as provided in Table \ref{table1}, and the estimated disaster-induced damages. Note that the predicted energy shortage is for a whole distribution network, and one period in Table \ref{table1} represents four hours. The cost and the size of available MERs are provided in Table \ref{table3}. For the Case Study, we have assumed that costs of MDGs, MESSs, and MPVs are equal, i.e., $C_j^{MDG}=C_j^{MESS}=C_j^{MPV}$, which implies that objective function \eqref{Preobj} minimizes the size of required MERs in the pre-disaster model. However, distribution system operators (DSOs) can choose these cost parameters based on their available markets. Post-disaster available DERs (PDA-DERs) such as DGs, ESSs, and PVs are also considered in the pre-disaster planning model. The availability of DERs in the immediate post-disaster restoration, as discussed before, is obtained using machine learning-based outage prediction models. The details of considered PDA-DERs in both Case Study I (CS-I) and Case Study II (CS-II) are given in Table \ref{table2}. For both Case Study, the fuel reserves for DGs and MDGs are assumed to be sufficient for the entire duration of restoration. The optimal mix of needed MERs for CS-I and CS-II, obtained from the pre-disaster planning model, is provided in Table \ref{table8}.

Then, the post-disaster restoration model is executed to determine the optimal locations of pre-determined MERs and the optimal operation period of damaged distribution lines. In addition to line outage information from post-disaster assessment and optimal mix of MERs obtained from the pre-disaster planning model, other parameters of the post-disaster restoration model are repair times of damaged lines (as in Table \ref{table5}) and the crew travel times in the network. Note that due to lack of relevant data, crew travel times in the networks were randomly generated in \textit{MATLAB} for both Case Study, and for the brevity, they are omitted in the paper. Also, note that available grid power for the entire restoration window is assumed to be zero for both Case Study; however, it can be easily included in the optimization if available. 

\vspace{-3mm}

\subsection{Case Study I}
As a result of the super-node simplification, the post-disaster IEEE 37-node feeder is reduced to seven aggregated nodes and six damaged lines in the optimization model. The operation period of damaged lines obtained from the post-disaster restoration model is provided in Table \ref{table7}, where one period represents four hours. The optimal locations of MERs obtained from the post-disaster model are provided in Table \ref{table8}. Note that two pre-determined RCs are used to repair damaged lines in this case study. As seen from the Table, two RCs repair three damaged lines each, all the damaged lines come to operation by period 7. 

\vspace{-3mm}

\subsection{Case Study II}
The super-node approximation simplifies the large-scale post-disaster IEEE 123-node feeder to nine super-nodes and eight damaged lines. Table \ref{table7} provides the operation period of damaged lines. Besides, the optimal locations of MERs are provided in Table \ref{table8}. It is noted that two pre-determined RCs are used in this case study as well. As seen from the Table, two RCs repair four damaged lines each, all the damaged lines are brought to operation by period 8.

In both Case Study, the obtained locations of MERs and the operation period of damaged lines are the optimal solutions for achieving the minimum unserved energy for the entire restoration window. Bar charts of the total energy required (TER) vs. the total energy supplied (TES) for each period are provided in figures \ref{ENS1} and \ref{ENS2} for CS-I and CS-II, respectively; note that one period represents four hours. The difference of TER and TES is unserved energy. It is worth noting that unserved energy gets smaller and smaller when more and more damaged lines come to operation. The operation period of damaged lines is affected by various system-specific parameters, such as availability of grid power, locations and presence of PDA-DERs, amount of ERRs, repair times of damaged lines, and crew travel times in the network.

\vspace{-3mm}

\subsection{Computational Performance}
The computational performance of proposed solution approaches, convex hull relaxation (CHR) and super-node approximation (SNA), was compared individually and jointly with MINLP form of the original problem, as shown in Table \ref{tabletime}. In Table \ref{tabletime}, \textit{Base} refers to the original MINLP form of the post-disaster restoration model without CHR and SNA. Similarly, \textit{SNA} refers to the MINLP form of the model with SNA, but without CHR. \textit{CHR} refers to the convex form of the model after CHR, but without SNA. \textit{SNA+CHR} refers to the convex form of the model with SNA and CHR. The NEOS server could not solve \textit{Base} and \textit{SNA} forms of the model, citing that computation exceeded the time limit of $8$ hours, which indicates that MINLP form of the model is computationally intractable. \textit{CHR} form of the model for CS-I and CS-II was solved in approximately $1$ hour and $6$ hours, respectively. Similarly, the \textit{SNA+CHR} form of the model took $33$ seconds and $228$ seconds for CS-I and CS-II, respectively. It is seen that \textit{CHR} and \textit{SNA+CHR} forms have almost identical objective values; however, the \textit{SNA+CHR} form of the model for both CS-I and CS-II was solved remarkably efficiently.


\section{Conclusion} \label{sec:conclusion}
In the presented paper, we have proposed a practical approach for disaster-resilient restoration of the distribution systems using ERRs in coordination with PDA-DERs. The handling of MERs from procurement/purchase to their operation in restoration is investigated in detail. The pre-disaster planning model is proposed to obtain the minimum amount of ERRs needed for a predicted post-disaster energy shortage, using outage prediction models. In the post-disaster restoration, damage assessment information and the pre-determined ERRs from the pre-disaster model are fed to the post-disaster model. The model co-optimizes ERRs and PDA-DERs to minimize the unserved energy, obtaining the optimal operation period of damaged distribution lines.

The proposed post-disaster model in MINLP form is convexified first using convex hull relaxation. Besides, the super-node approximation is introduced to solve the MICHP model for larger test feeders efficiently. The proposed approach and solution methods are tested using the IEEE 37 and 123 node test feeders. The test results indicate the successful pre-disaster and post-disaster allocation of ERRs in coordination with PDA-DERs for the disaster-resilient restoration of the distribution system.


\appendices
\section{Definition of Auxiliary Variables}
The definition of auxiliary variables for convex hull relaxation obtained in Section III-B is provided below:\\
$\boldsymbol{A}=diag(\left [\sqrt{2}, \sqrt{2}, 1, 1  \right ]^T)$,\\
$\boldsymbol{b}=\left [0 \; 0 \; 1 \; 1  \right ]^T$,\\
$\boldsymbol{x}_{ik,t}=\left [p_{ik,t} \; q_{ik,t} \; \ell _{ik,t} \; v_{i,t}\right]^T$, \\
$\boldsymbol{y}_{i,t}^{\alpha}=\left [p_{i,t}^{\alpha} \; q_{i,t}^{\alpha} \; p_{i,t}^{\alpha,l} \; v_{i,t}\right]^T$, \\
$\boldsymbol{A}_i^{\alpha}=diag(\left [0, \sqrt{2r_{i}^{\alpha,bt}}, 1, 1  \right ]^T)$,\\
$\boldsymbol{c}_{ik}=\left [0 \; 0 \; \underline{v}_{i}\overline{v}_{i} \; \left ( \overline{S}_{ik} \right )^{2}  \right ]^T$,\\
$\boldsymbol{c}^{\alpha}_{i}=\left [0 \; 0 \; \underline{v}_{i}\overline{v}_{i} \; r_{i}^{\alpha,e} \left ({S}_{i}^{\alpha\_t} \right )^{2}  \right ]^T$,\\
$d_{ik}=\left ( \overline{v}_{i} + \underline{v}_{i} \right )\left ( \overline{S}_{ik} \right )^{2}$,\\ $d_{i}^{\alpha}=r_{i}^{\alpha,e}\left ( \overline{v_{i}} + \underline{v_{i}} \right )\left ({S}_{i}^{\alpha\_t} \right )^{2}$,\\
$e_{i}^{\alpha}=r_{i}^{\alpha,e}\left ( {S}_{i}^{\alpha\_t} \right )^{2}$






\bibliographystyle{IEEEtran}
\bibliography{JournalRef.bib}

\begin{thebibliography}{10}
\providecommand{\url}[1]{#1}
\csname url@samestyle\endcsname
\providecommand{\newblock}{\relax}
\providecommand{\bibinfo}[2]{#2}
\providecommand{\BIBentrySTDinterwordspacing}{\spaceskip=0pt\relax}
\providecommand{\BIBentryALTinterwordstretchfactor}{4}
\providecommand{\BIBentryALTinterwordspacing}{\spaceskip=\fontdimen2\font plus
\BIBentryALTinterwordstretchfactor\fontdimen3\font minus
  \fontdimen4\font\relax}
\providecommand{\BIBforeignlanguage}[2]{{%
\expandafter\ifx\csname l@#1\endcsname\relax
\typeout{** WARNING: IEEEtran.bst: No hyphenation pattern has been}%
\typeout{** loaded for the language `#1'. Using the pattern for}%
\typeout{** the default language instead.}%
\else
\language=\csname l@#1\endcsname
\fi
#2}}
\providecommand{\BIBdecl}{\relax}
\BIBdecl

\bibitem{hines2008trends}
P.~Hines, J.~Apt, and S.~Talukdar, ``Trends in the history of large blackouts
  in the united states,'' in \emph{2008 IEEE Power and Energy Society General
  Meeting-Conversion and Delivery of Electrical Energy in the 21st
  Century}.\hskip 1em plus 0.5em minus 0.4em\relax IEEE, 2008, pp. 1--8.

\bibitem{executive2013economic}
E.~O. of~the President. Council~of Economic~Advisers, \emph{Economic Benefits
  of Increasing Electric Grid Resilience to Weather Outages}.\hskip 1em plus
  0.5em minus 0.4em\relax The Council, 2013.

\bibitem{smith2013us}
A.~B. Smith and R.~W. Katz, ``Us billion-dollar weather and climate disasters:
  data sources, trends, accuracy and biases,'' \emph{Natural hazards}, vol.~67,
  no.~2, pp. 387--410, 2013.

\bibitem{house2013presidential}
W.~House, ``Presidential policy directive--critical infrastructure security and
  resilience,'' \emph{Press Release, February}, vol.~12, 2013.

\bibitem{nagata2002multi}
T.~Nagata and H.~Sasaki, ``A multi-agent approach to power system
  restoration,'' \emph{IEEE transactions on power systems}, vol.~17, no.~2, pp.
  457--462, 2002.

\bibitem{adibi2000power}
M.~Adibi, ``Power system restoration,'' \emph{Methodologies and Implementation
  Strategies. IEEE series on Power Engineering. PM Anderson. Series Editor},
  2000.

\bibitem{arif2017networked}
A.~Arif and Z.~Wang, ``Networked microgrids for service restoration in
  resilient distribution systems,'' \emph{IET Generation, Transmission \&
  Distribution}, vol.~11, no.~14, pp. 3612--3619, 2017.

\bibitem{Chen2016}
C.~Chen, J.~Wang, F.~Qiu, and D.~Zhao, ``{Resilient Distribution System by
  Microgrids Formation after Natural Disasters},'' \emph{IEEE Transactions on
  Smart Grid}, vol.~7, no.~2, pp. 958--966, 2016.

\bibitem{kim2018enhancing}
J.~Kim and Y.~Dvorkin, ``Enhancing distribution system resilience with mobile
  energy storage and microgrids,'' \emph{IEEE Transactions on Smart Grid},
  2018.

\bibitem{lei2016mobile}
S.~Lei, J.~Wang, C.~Chen, and Y.~Hou, ``Mobile emergency generator
  pre-positioning and real-time allocation for resilient response to natural
  disasters,'' \emph{IEEE Transactions on Smart Grid}, vol.~9, no.~3, pp.
  2030--2041, 2016.

\bibitem{li2016convex}
Q.~Li, R.~Ayyanar, and V.~Vittal, ``Convex optimization for des planning and
  operation in radial distribution systems with high penetration of
  photovoltaic resources,'' \emph{IEEE Transactions on Sustainable Energy},
  vol.~7, no.~3, pp. 985--995, 2016.

\bibitem{sharma2019scenario}
S.~Sharma, Q.~Huang, A.~Tbaileh, and Q.~Li, ``Scenario-based analysis for
  disaster-resilient restoration of distribution systems,'' in \emph{2019 North
  American Power Symposium (NAPS)}.\hskip 1em plus 0.5em minus 0.4em\relax
  IEEE, 2019, pp. 1--6.

\bibitem{kabir2019predicting}
E.~Kabir, S.~D. Guikema, and S.~M. Quiring, ``Predicting thunderstorm-induced
  power outages to support utility restoration,'' \emph{IEEE Transactions on
  Power Systems}, vol.~34, no.~6, pp. 4370--4381, 2019.

\bibitem{guikema2010prestorm}
S.~D. Guikema, S.~M. Quiring, and S.-R. Han, ``Prestorm estimation of hurricane
  damage to electric power distribution systems,'' \emph{Risk Analysis: An
  International Journal}, vol.~30, no.~12, pp. 1744--1752, 2010.

\bibitem{nateghi2014forecasting}
R.~Nateghi, S.~D. Guikema, and S.~M. Quiring, ``Forecasting hurricane-induced
  power outage durations,'' \emph{Natural hazards}, vol.~74, no.~3, pp.
  1795--1811, 2014.

\bibitem{baran1989optimal}
M.~E. Baran and F.~F. Wu, ``Optimal capacitor placement on radial distribution
  systems,'' \emph{IEEE Transactions on power Delivery}, vol.~4, no.~1, pp.
  725--734, 1989.

\bibitem{li2017convex}
Q.~Li and V.~Vittal, ``Convex hull of the quadratic branch ac power flow
  equations and its application in radial distribution networks,'' \emph{IEEE
  Transactions on Power Systems}, vol.~33, no.~1, pp. 839--850, 2017.

\bibitem{li2018micro}
Q.~Li, S.~Yu, A.~S. Al-Sumaiti, and K.~Turitsyn, ``Micro water--energy nexus:
  Optimal demand-side management and quasi-convex hull relaxation,'' \emph{IEEE
  Transactions on Control of Network Systems}, vol.~6, no.~4, pp. 1313--1322,
  2018.

\bibitem{ulusan2018restoration}
A.~Ulusan and O.~Ergun, ``Restoration of services in disrupted infrastructure
  systems: A network science approach,'' \emph{PloS one}, vol.~13, no.~2, 2018.

\bibitem{hirschberg1979computing}
D.~S. Hirschberg, A.~K. Chandra, and D.~V. Sarwate, ``Computing connected
  components on parallel computers,'' \emph{Communications of the ACM},
  vol.~22, no.~8, pp. 461--464, 1979.

\end{thebibliography}



%







\end{document}